\documentclass[twocolumn,showpacs,aps]{revtex4}

\def\bea{\begin{eqnarray}}
\def\eea{\end{eqnarray}}
\def\ben{\begin{equation}}
\def\een{\end{equation}}
\def\benu{\begin{enumerate}}
\def\enu{\end{enumerate}}

\def\n{n}

\def\sss{\scriptscriptstyle\rm}

\def\1var{(\bx_1...\bx\N)}

\def\half{\frac{1}{2}}

\def\bp{{\bf p}}
\def\br{{\bf r}}

\def\bx{{x}}

\def\bq{{\bf q}}
\def\bj{{\bf j}}

\def\s{_{\sss S}}
\def\xc{_{\sss XC}}

\def\N{_{\sss N}}
\def\H{_{\sss H}}

\def\ext{_{\rm ext}}

\def\sph_int{ {\int d^3 r}}

\def\bA{{\bf A}}
\def\bE{{\bf E}}

\def\bP{{\bf P}}
\def\bp{{\bf p}}

\def\bG{{\bf G}}
\def\bq{{\bf q}}

\def\EM{_{\rm EM}}
\def\G{_{\rm G}}
\def\T{_{\rm T}}

\def\mac{^{\rm mac}}

\def\d2{\partial^\mu\partial_\mu}
\def\b0{{\bf 0}}

\begin{document} 

\preprint{RUTGERS DFT GROUP: pre-print MSB02}

\title{Current-Density Functional Theory of the Response of Solids}

\author{Neepa T. Maitra, Ivo Souza, and Kieron Burke}
\affiliation{Dept of Chemistry and Chemical Biology and Dept of
Physics and Astronomy, Rutgers
University, Piscataway, NJ 08854} 
\date{\today}


\begin{abstract} 
The response of an extended periodic system to a homogeneous field (of
 wave-vector $q=0$) cannot be obtained from a $q=0$ time-dependent
 density functional theory (TDDFT) calculation, because the
 Runge-Gross theorem does not apply.  Time-dependent {\em
 current}-density functional theory is needed and demonstrates that
 one key ingredient missing from TDDFT is the macroscopic current.  In
 the low-frequency limit, in certain cases, density polarization
 functional theory is recovered and a formally exact expression for
 the polarization functional is given.
\end{abstract}


\date{submitted: May 8, 2002; revised \today}
\maketitle

\vskip 3 cm
\def\section#1{}
Density functional theory\cite{HK64,KS65} is 
a standard approach for calculating ground-state
properties of solids\cite{JG89}
and molecules\cite{Kb99}.
Time-dependent
density functional theory (TDDFT) is an extension
of the ground-state formalism based on  the
Runge-Gross theorem \cite{RG84}; this establishes a one-to-one
correspondence between time-dependent densities
and time-dependent one-body potentials.
When a time-dependent electric field is applied
to a system, this formalism provides a route
to its optical response\cite{PGG96}.
The response equations of TDDFT have been
encoded in standard quantum chemical
packages\cite{C96}, and results for molecules are appearing
(see Ref. \cite{MBAG01} for many examples).
As in the ground-state case,
the accuracy depends on the quality
of the approximate functional used. 

There is great interest in applying the
same technique to extended systems.
While these can be treated well within existing
wavefunction technology, using, e.g., the GW
approximation and then solving the Bethe-Salpeter
equation for the optical response\cite{ARDO98}, the allure of
a TDDFT approach is its far lower computational cost.
Calculations already show
that excitonic effects appear to be treatable
by going beyond the usual local and semi-local approximations
of standard DFT calculations\cite{KG02,RORO02}.

There is also a version of the time-dependent theory, called
time-dependent current-density functional theory (TDCDFT), that uses
the current-density as the basic variable: As the choice of variable
(charge density versus current density) appears a matter of convenience,
TDCDFT and TDDFT appear to be equivalent (for non-magnetic
systems).  The time-dependent exchange-correlation potential has been
argued to be more amenable to local and semilocal approximation in
terms of the current-density~\cite{VK96} and this framework has been
used in recent response calculations of solids~\cite{KBS00,BKBL01} and
 conjugated polymers \cite{FBLB02}. Initial work towards a matrix
formulation of the current-density response equations has been
presented in Ref.~\cite{AZ02}.

In this paper, we demonstrate a difference {\em in principle}
between the two
approaches when applied to bulk solids.
The basic theorems of DFT, ground-state or time-dependent,
are proven for finite electronic systems (i.e. systems with a
boundary).  We consider the response of  periodic
systems (such as the bulk of an insulator or metal) to
time-varying electric
fields which have a 
spatially uniform component.  We show that
TDDFT fails in this case: there is no one-to-one correspondence between the
time-varying periodic density and the applied potential.  
Instead,  TDCDFT
is needed for a complete description.  
The time-dependent potential and all response properties
are {\em not} functionals of the time-dependent periodic density, but rather
{\em are} functionals of the time-dependent periodic current density.
In the limit of low frequencies
our analysis recovers the well-known GGG theorem
\cite{GGG95}, 
if microscopic transverse  currents can be neglected; TDCDFT 
then recovers (static) density polarization functional
theory \cite{GGG95,MOb97}.

Modern solid-state calculations 
model extended periodic systems and extract bulk properties 
by using periodic boundary conditions.  Our first point is that the
Runge-Gross (RG) theorem does not apply when a homogeneous electric
field is applied to a periodic system.  RG states that, given an
initial state, there is a one-to-one correspondence between
time-dependent densities and time-dependent scalar potentials for a given
interaction and statistics.  The first step of the RG proof
establishes a one-to-one correspondence between potentials and {\em
currents} \cite{XR85}.  In the second step, 
continuity is then used to relate currents to
densities, and a one-to-one mapping between densities and potentials
results provided a certain surface integral involving the 
density and the potential vanishes.
For
finite physical systems the condition for requiring this surface term 
to be zero can be
given rigorously for systems in which the density vanishes at the
surface~\cite{GK90}.  
 For a periodic system, one might try to choose a surface around which the density and potential are periodic but
 for a uniform field the linearity of the
 potential prevents this, and
 TDDFT does not apply.

Another way to see this is in modelling 
the periodic system in an electric field  by a large ring
 of length $L$ of the material, through which is threaded a
 time-varying solenoidal magnetic field \cite{K64}.  This field
 produces a uniform vector potential on the system, $\bA(t)$, that
 corresponds to a macroscopic electric field along the ring.  The
 beauty of this appoach is that the Hamiltonian remains spatially
 periodic, although it becomes time-dependent.  TDDFT was however derived with
 only scalar potentials in mind, and does not consider such transverse
 vector potentials; such uniform electric fields cannot be generated by 
a charge distribution. 
The first part of RG still holds in this case however, showing the
potential is a unique functional of the current-density.

Thus the density in the interior of any system
is insufficient information to deduce the external
electric field, 
but the current-density is sufficient.  This is a time-dependent
generalization of the original GGG theorem\cite{GGG95}.
We shall come back to the static case shortly.

A simple example demonstrating the non-uniqueness of the
potential-density mapping is a noninteracting free electron gas on
a ring, subjected to a constant, uniform electric field, $\cal E$,
turned on at time $t=0$.  Representing the field by a vector potential
$A = -c {\cal E} t $, each orbital in the Slater determinant state
satisfies the time-dependent Schr\"odinger equation
\ben 
(\hat p -{\cal E} t)^2\phi_m/2
=i\dot\phi_m 
\een 
(We use atomic units throughout this paper).  If the
electrons are initially in an eigenstate, $\phi_m(x,0) =
e^{i2\pi mx/L}/\sqrt{L}$, where $x$ is the coordinate around the
ring, and the conjugate momentum is $k_m = 2\pi m/L$, with
$m$ an integer between $0$ and $L$, different for each orbital, then
the solution at time $t$ is readily found to be \ben \phi_m(x,t)
=e^{-i\left(k_m^2 t/2 -k_m{\cal E}t^2/2 + {\cal E}^2t^3/6\right)} e^{i
2\pi m x/L}/\sqrt{L}\,.  \een Since the electric field only affects
the phase of this orbital, its density, and that of the
noninteracting gas, $n(x,t) = \sum_{m=1}^N
\vert\phi_m(x,t)\vert^2$, remains spatially uniform forever.
  In
particular, two different electric fields give rise to exactly the
same time-(in)dependent density. 
Thus the external potential is {\em
not} uniquely determined by the density here. (This argument holds for any number $N$ of electrons).

The first part of RG remains valid\cite{XR85} and applies
to arbitrary vector potentials, not just those
describing an electric field.
Choosing a  gauge in which
the scalar potential vanishes, one can show there is a one-to-one
correspondence between $\bA(\br t)$ and $\bj(\br t)$ for a given 
initial state \cite{GD88,N89,GDP96} and this provides the formal basis for TDCDFT.
In this gauge all electric fields are represented by  vector
 potentials but the one-to-one correspondence is of course gauge-independent.
In our simple example, 
 the physical current density is given by
$
j(t) =\sum_m^Nk_m +N{\cal E}t/L \,. 
$
In two different electric fields, two different currents arise.
More generally, for a periodic potential
on a (1-dimensional) ring,
when a uniform electric field is turned on, the density, current,
etc., remain periodic, and each can be written as 
$\sum j_G \exp(i G x)$, where $G=2n\pi/a$, and $n$ is an integer.
All components of the current-density at $G\neq 0$ are determined by the periodic density by the 
 continuity equation:
\ben
j_G (\omega) =  \omega n_G (\omega) /G,~~~~(G\neq 0).
\label{jG}
\een 
When a uniform electric field  is present, the $G=0$
component (the macroscopic current) is undetermined by the
time-dependent periodic density.

When the wavelength of the external field is finite, RG does apply to
 the periodic system; the macroscopic current is a functional of the
 periodic density (although cannot be determined by
 Eq.~(\ref{jG})). Examining the second step in the RG proof in
 one-dimension, one may choose the surface to be two points separated
 by an integral multiple of the lattice constant and the wavelength of
 the external field: the Hamiltonian is periodic around this surface,
 and the integrand of the surface term is the same at the two points.
 Thus the surface term vanishes and the RG theorem holds.  This is
 consistent with the ground-state case \cite{MOb97} where the
 macroscopic polarization is only an independent variable when the
 external field has a homogeneous component. In our simple example,
 consider applying a perturbative electric field ${\cal E}\cos(qx)$
 around the ring, where a finite number of wavelengths fits into the
 ring so $q$ is an integer multiple of $2\pi/L$.  The density picks up
 a spatial modulation, which, in the limit that $q\to 0$, becomes, to
 first order in ${\cal E}$, \ben n(xt) \to N\left(1 +{\cal E} q
 \sum_{m=1}^N (x+k_m t)^3/3\right)/L .  \een Different fields yield
 different densities except when $q=0$, consistent with the 1-1
 mapping between densities and potentials at finite wavelengths.  One
 can then imagine attempting to find the uniform-field value of
 certain response functions \cite{AK60} by taking the $q\to 0$ limit
 of a series of finite-$q$ TDDFT calculations, although the larger
 supercell required might render this procedure impractical (although, see discussion of Ref.~\cite{KG02} below). The same
 issue arises in the ground-state case, where it was shown in
 Ref. \cite{KR83} how careful use of a sawtooth potential on a
 supercell can resolve this problem. The approximation for the
 density-functional must be ultra-nonlocal in space in order to
 capture the exchange-correlation fields generated from the charge
 distribution modulated by the long wavelength. The first excitonic
 peak in the dielectric function of silicon (missing when using any local 
or semi-local approximation in TDDFT) 
 was captured in
 Ref.~\cite{RORO02} in a TDDFT calculation at finite wavevector, using
 an ultranonlocal kernel $f\xc \sim\alpha/q^2$ however with $\alpha$
 being empirically determined, and with applying a GW shift to the
  spectrum. In Ref.~\cite{KG02} an exact exchange calculation was
 performed, reproducing the excitonic peak of silicon, 
without having to battle through a monstrously large numerical computation. 
 The authors
 were able to analytically extract the diverging ultranonlocal factor
 $1/q^2$ from the exact-exchange kernel and so needed to numerically evaluate 
only the
 remaining matrix elements, well-behaved as a function of $q$: because the long-range behaviour had been  already factored out,
 the numerical computation converged rapidly with the number of
 $k$-points.

 However, like density-polarization
 functional theory in the ground-state case, TDCDFT allows one to obtain
 the homogeneous field result directly from a single $q=0$
 calculation; this cannot be done in TDDFT.

Having established that the current is needed
 when calculating bulk response
properties
in solids in uniform fields, we next
review how such a calculation is performed for any electronic
system.  
The one-to-one correspondence between $\bA(\br t)$ and $\bj(\br t)$
 can be used to establish a set of KS equations
in which noninteracting electrons move in a KS vector
potential $\bA\s(\br t)$ and reproduce the exact current-density $\bj(\br t)$.
 Assume the exact KS ground-state has been found and
is nondegenerate.  The TDCDFT KS equations are
\ben
\left\{\half \left[\bp + \frac{1}{c}\bA\s[\bj](\br t)\right]^2 + v\s(\br)\right\}
\phi_i(\br t) = i \frac{\partial \phi_i(\br t)}{\partial t}
\label{KSeqn}
\een
where $v\s(\br)$ is the periodic ground-state KS potential 
and all time-dependence is in the KS vector potential 
$\bA\s[\bj](\br t)$.  
The orbitals begin as the occupied ground-state KS orbitals.
The KS vector potential is defined to produce the
exact physical current density 
\ben
\bj(\br t) = \Im \sum_{i occ} \phi_i^*(\br t) \nabla \phi_i(\br t)
-\frac{1}{c}\n(\br t) \bA\s(\br t)\,.
\label{j}
\een
Departing from convention, we write
\ben
\bA\s(\br t)=\bA\ext(\br t)+ 
\bA\EM (\br t) + \bA\xc (\br t).
\label{eq:As}
\een
Here $\bA\EM (\br t) = \bA\H (\br t) +\bA\T(\br t)$ is the full electromagnetic
potential, satisfying Maxwell's equation:
\ben
\left\{\nabla^2 - \frac{1}{c^2} \frac{\partial^2}{\partial t^2} \right\}
\bA\EM -\nabla\cdot\left(\nabla\cdot\bA\EM\right)=
-\frac{4\pi}{c}\bj \,.
\een
The longitudinal component, $\bA\H$, is the vector
equivalent of the Hartree potential, while 
the transverse component, $\bA\T$, arises from the transverse
component of the current:
\ben
\bA\T(\br t)=
\frac{1}{c}\int d^3r'\ 
\frac{\bj\T(\br,t-\vert\br-\br'\vert/c)}{\vert\br - \br'\vert} \,.
\label{ATdef}
\een In previous formulations of TDCDFT (e.g., Ref. \cite{VK96}), this
term does not appear explicitly. 
 Since $\bA\T$ is a
nonlocal classical electromagnetic contribution of any moving charge
density, it is {\em not} an exchange-correlation effect, and should be
included {\em exactly} in any time-dependent calculation.  
Unlike the
Hartree contribution, it is not adiabatic, i.e., it depends on the
retarded current, not the instantaneous density.  Consider the one-dimensional
example above cast in three dimensions: we fatten the ring in
the radial and cylindrical directions, imposing either hard wall or
periodic boundary conditions in these two directions.  The current
flows only azimuthally and uniformly on the ring and so is purely
transverse in the sense that it has nonzero curl and zero
divergence. At the same time, we call it macroscopic, since it is
uniform along the ring. The classical response appears purely from 
$\bA\T$ rather than from Hartree. We have checked that the
 VK approximation  for the exchange-correlation vector potential \cite{VK96}
 is unaffected by the addition of $\bA\T$.  
(We note that the addition of $\bA\T$ to static current-density
functional theory has been discussed for example in \cite{VR87}.)

We pause to make a connection with TDDFT. 
The TD density is given exactly in TDCDFT, via
continuity 
$\dot n(\br t) = - \nabla \cdot \bj(\br t)$
but a 
TDDFT calculation is only guaranteed to reproduce the longitudinal
component of the current.
If there does exist a TDDFT KS potential
$v\s(\br t)$ that reproduces the interacting current as well as the
density\cite{MBAG01}, then the TDCDFT potential is  
$\bA\s(\br t) = c\int^tdt'\nabla v\s(\br t')$.
But there exist special geometries, like the ring, to which
TDDFT cannot be applied at all when macroscopic fields are present.

Next we consider the special case of linear 
response to a uniform electric field,
$A\ext(\br t)=c{\cal E}(\omega)\exp(-i\omega t +\eta t)/i\omega$, 
where $\eta=0_+$
turns the field on adiabatically from $t=-\infty$ to $0$.
If the current does not grow indefinitely (so that Fourier
transforms exist), 
all additions to the external potential in Eq. (\ref{eq:As}) can
be written to first order in the perturbation:
\bea
\nonumber
\bA\H(\br\omega)&=&
\frac{c}{(i\omega)^2}\int d^3r'\nabla\frac{1}
{\vert \br-\br'\vert}\nabla'\cdot\bj(\br'\omega)
\nonumber\\
\bA\T(\br\omega)&=&
\frac{1}{c}\int d^3r'\frac{\exp(i\omega (\vert\br-\br'\vert/c))
\bj\T(\br'\omega)}
{\vert\br - \br'\vert}
\nonumber\\
\bA\xc(\br \omega)&= &\int d^3r'\
\tensor{f}\xc[\n_0](\br\br'\omega)\cdot\
\bj (\br'\omega)
\label{fxcdef}
\eea
where $\tensor{f}\xc$ is a nonlocal exchange-correlation
tensor functional of the ground-state density,
analogous to the scalar exchange-correlation kernel of TDDFT.

We return now to solids in time-varying electric fields, and 
consider the macroscopic
response, using the ring geometry as described earlier. 
Let $\bG$ be the reciprocal lattice vector. The density change when the external field is turned on
\ben
\delta \n(\br t) = \sum_{\bG\neq 0} \delta \n_\bG(\omega) \exp(i\bG\cdot\br
-i \omega t)
\een
has no macroscopic (i.e., $\bG=0$) component due to charge conservation. 
 This 
implies that the Hartree response remains always periodic
and has no macroscopic component.
Performing a spatial
Fourier transform on the transverse potential yields
\ben
\bA_{{\rm T}\bG}(\omega)=
-\frac{4\pi c}{\omega^2 - c^2 {\rm G}^2}\ \bj_{{\rm T}\bG}(\omega).
\een
We distinguish the microscopic ($\bG\neq 0$) transverse
current from the macroscopic
($\bG=0$) current, which is a
spatially uniform current
travelling
around the ring.
As $\omega\to 0$, the microscopic
contribution vanishes relative to Hartree, because
of the factor of $\omega^2$ in Eq. (\ref{fxcdef}).
The macroscopic component, however,  does not, yielding
\bea
\bA\T\mac (\omega)&=&\frac{4\pi c}{(i\omega)^2}\
\bj_{\bG=0} (\omega)
\nonumber\\
\bA\xc\mac(\omega)&=&\sum_{\bG} 
\tensor{f}_{{\sss XC}0\bG}[\n_0](\omega)\cdot\
\bj_\bG(\omega).
\label{Axcmac}
\eea
Thus, from the ring perspective, the origin of the electromagnetic response
(traditionally considered a Hartree effect) is the {\em transverse}
potential generated by the ring current.  In the limit
when $L\to\infty$, the distinction between transverse and
longitudinal breaks down, but for any
real calculation with discrete $k$-points in the Brillouin zone, 
this distinction is important. 
For any $L$, $\bA\T\mac (\omega)$, being the classical 
macroscopic response, cannot be neglected in considering solids in electric 
fields.  

For insulators and metals at finite frequencies, 
after the adiabatic turn-on of the field, the system
settles into a steady state, thanks to
the damping factor $\eta$. Because we are using TDCDFT,
the macroscopic current in this KS calculation is the
physical one at all times.  Thus the macroscopic polarization
\ben
\bP\mac(t)=-\int^t_{-\infty} dt'\, \bj_{\bG=0}(t'),~~~
\bP\mac(\omega)=\frac{\bj_{\bG=0}(\omega)}{i\omega}
\label{Pmac}
\een
equals the physical one.  Multiplying the vector potential
by $i\omega/c$ yields the KS electric field:
\bea
\nonumber
\bE\xc\mac(\omega)&=&-\frac{\omega^2}{c}\tensor{f}_{{\sss XC}00}(\omega)
\bP\mac(\omega)
+\frac{i\omega}{c}\sum_{\bG\neq0} \\
&&\tensor{f}_{{\sss XC}0\bG}(\omega)\cdot\left\{\frac{\bG \omega \n\G(\omega)}
{{\rm G}^2}+\bj_{{\rm T}\bG}(\omega) \right\}
\label{Excmac}
\eea 
where we have used continuity.
If the microscopic transverse
currents vanish or can be neglected, then $\bj_\bG$ can be written solely
in terms of $\n_\bG$, meaning that the entire corrections to the
external field can be given in terms of the macroscopic polarization
and the periodic density change.  Now consider the final state of the
system after the field has been adiabatically switched on. Then in the
limit that $\omega\to 0$, this second point is precisely the GGG
assertion made in density polarization functional theory: that the
exchange-correlation field depends on both the periodic density and
the macroscopic polarization \cite{GGG95,MOb97}.  That result arises here
out of TDCDFT, even at finite frequencies, and Eq. (\ref{Excmac}) then
 provides an exact expression for the polarization functional. 
 However if microscopic
transverse currents cannot be neglected, then for the GGG result to
hold, these currents themselves would need to be functionals of the
periodic density and macroscopic polarization; this remains to be
investigated.

During the turning-on of the electric field, in the $\omega\to 0$
limit the system remains always in its adiabatic ground-state.  Thus
minimization of the energy (including the
exchange-correlation electric field terms) over periodic functions
\cite{NV94,MOb97,NG01}, within the Berry-phase formalism \cite{KV93,NV94},
will yield the same result as the TDCDFT calculation in the
low-frequency limit.

This work also shows that polarization is an infinite memory effect
within TDCDFT \cite{MBW02}, i.e., after the electric field has reached a
finite value, the system `remembers' forever the current that flowed
in turning it on. (It does not depend on the procedure in reaching the
steady-state, rather it depends just on the time-integrated current).
Similarly, in pumping a finite system from the ground-state to a given
excited state, the current provides a natural way for the system to
remember indefinitely which state it is in.  Thus TDCDFT may provide a
natural solution to some of the paradoxes generated by a pure
time-dependent DFT\cite{MBW02}.


Lastly, we discuss existing approximations and calculations.  Just
like the Hartree term, any {\em density} functional approximation,
e.g. ALDA or AGGA, misses entirely the macroscopic contributions
discussed here\cite{BIRY00}.  Even if we regard the bulk insulator as
a large but finite slab, so that TDDFT does in principle apply, we would
need an ultra-nonlocal functional of the density in order to capture
the contribution to the XC field from the polarization charge density
on the surfaces. In the static limit, this is (often) essentially the
GGG effect discussed above, where in addition to the bulk periodic
density, the functional depends on the bulk macroscopic
polarization.
For example,  for a large rectangular slab 
with translation symmetry, the exchange-correlation electric field is 
\ben
\bE\xc(\bq, \omega) = -i\bq f\xc(\bq, \omega) n^{(1)}(\bq,\omega)
\een
where the TDDFT kernel in the long wavelength limit
may be expressed in terms of the component of the TDCDFT kernel parallel
 to the field: $\lim_{\bq\to 0} f\xc(\omega) = -\omega^2{\tensor f}_{{\sss XC}00\parallel}(\omega)/(cq^2).$
  The ultranonlocal $1/q^2$ dependence multiplies the response
density, $n^{(1)}$: this has a lattice-periodic part in the bulk as
well as a part proportional to the polarization charge density on the
{\it surface}.  The fact that one needs to look only at a unit cell in
the bulk and integrate up the current that has flowed through it to
obtain $\bP\mac(t)$, shows how local approximations in terms of the
current capture the essential physics, while local density
approximations cannot. Since orbitals depend ultranonlocally on space,
orbital functionals in TDDFT \cite{UGG95} may capture polarization effects.
However, since polarization can be obtained from the current-density,
these more complex functionals are not necessary to describe these effects if
the current-density is used as the basic variable.


TDCDFT calculations for the optical response
of solids have already been reported
\cite{KBS00,BKBL01} that explicitly include the macroscopic transverse
component of $\bA\EM$ on top of the VK approximation.  
In adding the VK correction to their calculation\cite{VK96},
the microscopic contributions in Eq. (\ref{Excmac}) are
not included (Eq. (16) of Ref. \cite{BKBL01}).
This may be related to the need
to divide the beyond-ALDA contribution by a factor of 2.5.
Similarly, the simple approximation in density polarization functional theory
of Ref. \cite{AJW96},
in which $\bE\xc = \gamma \bP\mac$, where $\gamma$
is a constant, does not include such contributions.

We conclude by noting that while the present work has focussed on the
special case of the interior of a bulk insulator or metal, the underlying logic
was motivated by the need to construct an approximately local theory
of exchange-correlation for any system in an electric field.  Two
examples make this clear.  In tunnelling through a quantum wire,
present calculations use {\em ground-state} DFT KS orbitals as their
starting point.  This leads to resonances at the positions of bare KS
orbitals.  For finite systems without currents, regular DFT tells us
there are significant exchange-correlation corrections.  The only way
to calculate such corrections for a quantum wire is using TDCDFT, in
order to handle currents.  In another area, electron and energy
transfer in biological molecules, attempts are being made to estimate
matrix elements in a TDDFT calculation\cite{CBC02}.  Such calculations,
using adiabatic local and gradient-corrected approximations, clearly
miss any contributions from macroscopic currents.

This work was inspired by that of Profs. Kohn and
Tolkien, and
supported by the Office of Naval Research under
grant no. NOOOO14-01-1-1061.
We thank David
Vanderbilt, Ralph Gebauer, Roberto Car, Hardy Gross, and Giovanni Vignale
 for useful discussions.


\end{document}